\documentclass[preprint]{ptephy_v1}

\preprintnumber{KUNS-2787} 

\usepackage{graphicx}
\usepackage{epsf}
\usepackage{amsmath,amssymb}

\usepackage{float}
\usepackage{color} 
\usepackage{multirow}
\usepackage{dcolumn}
\usepackage{array}
\usepackage{comment}

\usepackage{tikz}
\usepackage{array,booktabs}

\usepackage{ulem}

\def\vector#1{\mbox{\boldmath $#1$}}
\def\hana#1{\mathcal{#1}}
\def\Hesix{{}^6\textrm{He}}
\def\Beten{{}^{10}\textrm{Be}}
\def\Carbon{{}^{12}\textrm{C}}
\def\Oxy{{}^{16}\textrm{O}}
\def\2-b{\alpha + \Hesix}
\def\3-b{\alpha + \alpha + 2n}
\def\bg{\beta\gamma}
\def\IntGS{0_\textrm{gs}^+}
\def\IntNDzero{1_\textrm{ND}^-(K=0)}
\def\IntNDone{1_\textrm{ND}^-(K=1)}
\def\IntCL{1_\textrm{cl}^-}
\usepackage[utf8]{inputenc}
\usepackage{natbib}
\begin{document}
\title{Variation after $K$-projection in antisymmetrized molecular dynamics  for low-energy dipole excitations in $^{10}$Be and $^{16}$O
}
\author{Yuki Shikata}
\author{Yoshiko Kanada-En'yo}
\affil{Department of Physics, Kyoto University, Kyoto 606-8502, Japan}


\begin{abstract}

For study of dipole excitations, a new method of variation after $K$-projection in the framework of
antisymmetrized molecular dynamics (AMD) with the deformation $\beta$ constraint was 
proposed. The method was applied to $^{10}$Be and $^{16}$O to describe low-energy dipole excitations and  found to be a useful and economical approach for dipole excitations.
In the application to $\Beten$, two dipole states in low-energy region were described.
For $\Oxy$, the $1_1^-$ and $1_2^-$ states with remarkable dipole strengths were obtained. 
The $1_1^-$ state is characterized by significant toroidal dipole (TD) and compressional dipole strengths, 
whereas  the $1_2^-$ has significant TD strength and shows a developed $\alpha + \Carbon$ cluster structure.
Dipole properties in $^{16}$O were discussed by analysis of current densities of the dipole transitions.  

\end{abstract}

\subjectindex{D11, D13}

\maketitle

\section{Introduction}

In recent studies of nuclear excitations in stable and unstable nuclei, 
low-energy dipole (LED) strength, which has been found in lower energy regions than giant dipole resonances, 
has attracted experimental~\cite{1402-4896-2013-T152-014012,Bracco:2015hca,Bracco:2019gza} and theoretical~\cite{Paar:2007bk} attention.
The LED strengths has been observed in some $N=Z$ nuclei~\cite{Harakeh:1981zz,Poelhekken:1992gvp,Youngblood:1999zz,John:2003ke} and in various neutron-rich nuclei such as $^{18}$O~\cite{PhysRevC.43.2127,Manley:1991zz}, $^{20}$O~\cite{Nakatsuka:2017dhs}, and $^{26}$Ne~\cite{Gibelin:2007fda,Gibelin:2008zz}.
Properties and origins of the LED excitations have been often discussed in relation with symmetry energy of nuclear matter~\cite{Brown:2000pd,Piekarewicz2014,Colo2014}, neutron skin thickness in neutron-rich nuclei~\cite{Piekarewicz:2006ip,Inakura:2011mv,Piekarewicz:2012pp,Tamii:2011pv,Birkhan:2016qkr}, and neutron capture rate from astrophysical interests~\cite{Goriely:1998utv,Goriely:2004qb,Tonchev:2017ily}, but they have been not clarified yet.

In theoretical studies of the LED excitations, a couple of kinds of LED modes have been proposed.
For example, the Pygmy mode in $N> Z$ nuclei, in which excess neutrons oscillate in an opposite direction to 
the core~\cite{Ikeda:pygmy}, is considered to contribute isovector dipole (IVD) strengths ~\cite{Mohan:1971tz,10.1143/PTP.83.180,VanIsacker:1992zz}.
Another type is toroidal dipole (TD) mode characterized by a vortical nature suggested by Refs.~\cite{Dubovik:toroidal,Semenko:toroidal}.
The toroidal mode can produce isoscalar dipole (ISD) strength and is expected to appear in both of neutron-rich and $N=Z$ nuclei.
In the recent years, the TD mode is intensively studied by microscopic models~\cite{Vretenar:2001te,Ryezayeva:2002zz,0954-3899-29-4-312,Reinhard:2013xqa}.
Kvasil {\it et al.} have introduced the TD operator to measure the toroidal feature 
and demonstrated that it is a sensitive probe for the TD mode~\cite{0954-3899-29-4-312}
though experimental observation of the TD strength has not been confirmed yet.
The third candidate is cluster excitation mode.
As pointed out by Chiba {\it et al.}~\cite{Chiba:2015khu}, the ISD strength can be enhanced in the cluster excitation mode. 
In the low-energy spectra, these different kinds of dipole modes do not necessarily appear as identical states but may mix with each other in the low-energy $1^-$ spectra.

In deformed nuclei, further interesting features of the LED excitations can be found due to
coupling of the dipole modes with nuclear deformation.
In a strong coupling regime, the dipole excitations can be classified by $K$ quanta, which indicates 
the $z$-component of total angular momentum in a body-fixed frame.  
Nesterenko {\it et al.} investigated the LED excitations in deformed nuclei with a mean field approach
\cite{Kvasil:2013yca,Nesterenko:2016qiw,Nesterenko:2017rcc}, and showed that compressional dipole and TD modes appear as 
$K=0$ and $K=1$ states, respectively.
The LED excitations in a deformed system of $\Beten$ have been also discussed in our 
previous work~\cite{Shikata:2019wdx}, in which properties of LED have been discussed in terms of 
$K$ quanta in the prolately deformed system. 

Our future goal is to investigate LED excitations in various nuclei including stable and unstable nuclei
with a microscopic framework and to reveal their fundamental features and origins.
For this aim, it is essential to describe various LED modes such as 
the Pigmy, TD, and cluster excitation modes in general nuclei with and without deformations.

In this paper, we propose a new method of variation after $K$-projection in a framework of the 
$\beta$-constraint antisymmetrized molecular dynamics (AMD) for the study of LED excitations. 
By using the AMD wave function, 
the energy variation is performed after the parity- and $K$-projections under constraint of the 
quadrupole deformation.
The AMD is a useful microscopic approach for structure study of light nuclei, which can describe developed cluster structures, shell-model-like states, and their intermediate states ~\cite{Kanada-Enyo:2001yji,KANADAENYO2003497,Kimura:2016fce}.
For study of deformed nuclei, the AMD has been developed to the $\beta$-constraint and $\beta\gamma$-constraint versions, 
in which quadrupole deformation parameters $\beta$ and $(\beta,\gamma)$ are constrained, respectively, in the energy variation~\cite{Dote:1997zz,10.1143/PTP.106.1153,Suhara:2009jb}.
The latter, $\beta\gamma$-constraint AMD, is a better approach than the former $\beta$-constraint
for detailed description of deformations, in particular, axial asymmetric structures. 
However, it needs numerical cost because of a huge number of basis wave functions in the two dimensional space of the constraint parameters
and may encounter difficulty in application to heavy-mass nuclei.  
In the present method, we improved the $\beta$-constraint by performing the $K$-projection in the energy variation to 
efficiently obtain basis wave functions essential to LED excitations. 

As a test case of a light deformed nuclei, we first apply the present method to $\Beten$.
Structure of $\Beten$ has been investigated by 
many theoretical studies, e.g., cluster model~\cite{Fujimura:1999zz,Descouvemont:2002mnw}, molecular orbital model~\cite{vonOertzen1996,Itagaki:2000nn}, and AMD~\cite{Kanada-Enyo:1999bsw,Kanada-Enyo:2015knx},
which show $2\alpha+nn$ cluster structures of low-energy states.
We compare results of the $\beta$-constraint AMD with and without the $K$-projection as well as the 
$\beta\gamma$-constraint AMD and show the applicability of the $K$-projection method. 
Then we apply the method to study dipole excitations of $\Oxy$. 

This article is organized as follows.
In Sect.\ \ref{sec:formalism}, the present formalism of $K$-projection after variation in the AMD framework is explained.
The generator coordinate method (GCM) and expression of dipole operators are also described.
The interactions used in the present calculation are expressed in Sect.\ \ref{sec:effective_int}.
Section\ \ref{sec:10Be} shows application to $\Beten$.
In Sect.\ \ref{sec:16O}, the LED excitations in $\Oxy$ are investigated with the present method.
Finally, this article is summarized in Sect.\ \ref{sec:summary}.

\section{Formalism} \label{sec:formalism}

In order to investigate the dipole excitations, 
we propose a new method of a constraint AMD with variation 
after the $K$-projection as well as the parity projection. That is,  
energy variation is performed for the AMD wave function projected 
onto the parity and the $z$ component of angular momentum eigen state
under the quadrupole constraint ($\beta$ or $\beta\gamma$).
After the energy variation, thus obtained AMD wave functions are superposed with GCM. For details of the AMD framework, 
the reader is referred to Refs.~\cite{Kanada-Enyo:1998onp, Kanada-Enyo:2001yji, 10.1143/PTP.106.1153}.

\subsection{AMD}
An AMD wave function is given by a slater determinant:
\begin{eqnarray}
\Phi = \mathcal{A}\left[\psi_1\psi_2\cdots\psi_A\right],
\end{eqnarray}
where $\psi_i\ (i=1,2,\cdots,A)$ is the $i$th single-particle wave function given by a localized Gaussian wave packet as
\begin{eqnarray}
\psi_i\ &=& \phi(\vector{Z}_i)\chi(\vector{\xi}_i)\tau_i, \\
\phi(\vector{Z}_i) &=& \left(\frac{2\nu}{\pi}\right)^{\frac{3}{4}}\exp\left[ -\nu\left(\vector{r} - \frac{\vector{Z}_i}{\sqrt{\nu}}\right)^2 \right],\\
\chi(\vector{\xi}_i)&=&\xi_{i\uparrow}|\uparrow\rangle + \xi_{i\downarrow}|\downarrow\rangle, \\
\tau_i &=&p\ \textrm{or}\ n.
\end{eqnarray}
$\vector{Z}_i$ and $\vector{\xi}_i$ indicate Gaussian centroids and nucleon-spin orientations, respectively, and they are treated independently as variational parameters determined by the energy optimization.
$\nu$ is the width parameter and is taken to be common with all nucleons.

In a simple version of AMD, the energy variation,
\begin{eqnarray}
\delta\left(\frac{\langle\Psi|\hat{H}|\Psi\rangle}{\langle\Psi|\Psi\rangle} \right) = 0 \label{eq:var}
\end{eqnarray}
is performed for the parity-projected AMD wave function $|\Psi\rangle = \hat{P}^{\pi}|\Phi\rangle$ with the parity projection operator $\hat{P}^{\pi}$.

\subsection{Quadrupole constraint AMD}
In order to perform the $K$-projection, energy variation is performed under constraint of the quadrupole deformation parameter $\beta$ or $\beta\gamma$.
The deformation parameters $\beta$ and $\gamma$ are defined as
\begin{eqnarray}
\beta\cos\gamma &=& \frac{\sqrt{5}}{3}\frac{2\langle z^2\rangle - \langle x^2\rangle - \langle y^2\rangle}{R^2}, \\
\beta\sin\gamma &=& \sqrt{\frac{5}{3}}\frac{\langle x^2\rangle - \langle y^2\rangle}{R^2}, \\
R^2 &=& \frac{5}{3}(\langle x^2\rangle + \langle y^2\rangle + \langle z^2\rangle),
\end{eqnarray}
where $\langle \vector{r} \rangle$ represents an expectation value of operator $\hat{\vector{r}}$ by the AMD wave function without any projection.

In this paper, we perform two kinds of constraint AMD.
One is $\bg$-constraint AMD($\beta\gamma$-AMD)~\cite{Suhara:2009jb} and the other is the $\beta$-constraint AMD($\beta$-AMD)~\cite{10.1143/PTP.106.1153}.
For a given constraint value $\beta_0$ (or values ($\beta_0,\gamma_0$)), the energy variation is done by 
imposing the constraint $\beta=\beta_0$ ($\beta=\beta_0,\gamma=\gamma_0$). 
Here we keep the condition for the 
off-diagonal components of moment of inertia as $\langle xy\rangle=\langle yz\rangle=\langle zx\rangle=0$.
After the energy variation, the optimized AMD wave function with the given quadrupole deformation 
$|\Phi(\beta_0)\rangle$ ($|\Phi(\beta_0,\gamma_0)\rangle$)  is obtained.

\subsection{K-projection VAP}
In the present method, the energy variation in eq.\ \eqref{eq:var} is performed for the $K$ and parity projected AMD wave function $|\Psi\rangle=\hat{P}_{K}\hat{P}^{\pi}|\Phi\rangle$ instead of $|\Psi\rangle=\hat{P}^{\pi}|\Phi\rangle$ without the $K$ projection.
Here,
$\hat{P}_K$ is the $K$-projection operator given as 
\begin{eqnarray}
\hat{P}_K = \frac{1}{2\pi}\int_0^{2\pi}d\theta\ e^{-iK\theta}\hat{R}(\theta),
\end{eqnarray}
where $\hat{R}(\theta)$ is the rotation operator around the principal axis in body-fixed frame.
We call the present method "K-VAP" and the conventional method "P-VAP" to distinguish the two methods.
As a result of K-VAP, it is expected that the AMD wave function can be optimized for the given parity and $K$ quantum number.
In order to describe the ground states wave functions of $\Beten$ and $\Oxy$,  
the $K^\pi=0^+$ projection is performed in K-VAP, and the
resultant AMD wave function is called $K^{\pi}=0^+$ basis.
For the dipole excitations, i.e., the $1^-$ excitation states, $K^\pi=0^-$ and $K^\pi=1^-$ projections are performed, and the obtained AMD wave functions are called $K^{\pi}=0^-$ and $K^{\pi}=1^-$ bases, respectively.
The obtained AMD wave function of $\beta$-AMD($\bg$-AMD) with K-VAP is denoted as 
$|\Phi_K^\pi(\beta)\rangle$($|\Phi_K^\pi(\beta,\gamma)\rangle$).

\subsection{GCM}
In order to construct wave functions for the ground and $1^-$ states, 
GCM is applied for $\beta$-AMD with K-VAP with respect to the  generator coordinate $\beta$. 
It means that the basis wave functions are superposed as
\begin{eqnarray}
\Psi^{\pi}(J_m) = \sum_{K,K'}\sum_{\beta}c_{KK'}(\beta)\hat{P}_{MK'}^J\hat{P}^{\pi}|\Phi_K^{\pi}(\beta)\rangle ,
\end{eqnarray}
where the parity projection and total angular momentum projection $\hat{P}_{MK'}^J$ are performed.
Note that, for the $1^-$ states, $K=0$ and $K=1$ bases are adopted and mixing of $K'$ components is taken into account.
The coefficients $c_{KK'}(\beta)$ are determined by solving the Hill-Wheeler equation~\cite{Hill:1952jb,Griffin:1957zza}.
 In a similar way, the GCM calculation with the P-VAP bases wave functions, $|\Phi^{\pi}(\beta_0)\rangle$, 
is also performed for comparison.
For the width parameter, we set $\nu = 0.235\ \textrm{fm}^{-2}$ for $\Beten$ and $\nu = 0.19\ \textrm{fm}^{-2}$ for $\Oxy$ which are same in Refs.~\cite{Shikata:2019wdx, Kanada-Enyo:2019hrm}.

\subsection{dipole operator}
To analyze the LED excitation, we use the three dipole operators:
$E1$, compressive dipole (CD), and toroidal dipole (TD) operators~\cite{0954-3899-29-4-312} defined as
\begin{eqnarray}
\hat{M}_{E1}(\mu) &=& \frac{N}{A}\sum_{i\in p}r_iY_{1\mu}(\hat{\vector{r}}_{i}) - \frac{Z}{A}\sum_{i\in n}r_iY_{1\mu}(\hat{\vector{r}}_{i}), \\
\hat{M}_{\textrm{CD}}(\mu) &=& \frac{-1}{10\sqrt{2}c}\int d\vector{r}\ \nabla\cdot\vector{j}_{\textrm{nucl}}(\vector{r})\ r^3Y_{1\mu}(\hat{\vector{r}}), \label{eq:CD_op}\\
\hat{M}_{\textrm{TD}}(\mu)&=&\frac{-1}{10\sqrt{2}c} \int d\vector{r}\ (\hat{\nabla}\times\vector{j}_{\textrm{nucl}}(\vector{r}))\cdot r^3\vector{Y}_{11\mu}(\hat{\vector{r}}), \label{eq:TD_op}
\end{eqnarray}
where $\vector{j}_{\textrm{nucl}}$ is convection nuclear current and $\vector{Y}_{jL\mu}(\hat{\vector{r}})$ is vector spherical given as
\begin{eqnarray}
\vector{j}_{\textrm{nucl}}(\vector{r}) &=& \frac{-i\hbar}{2m}\sum_{k=1}^A\{ \vector{\nabla}_k\delta(\vector{r}-\vector{r}_k) + \delta(\vector{r}-\vector{r}_k)\vector{\nabla}_k \},  \label{eq:current} \\
\vector{Y}_{jL\mu}(\hat{\vector{r}}) &=& \sum_{\alpha,\beta}\langle L\alpha,1\beta|j\mu\rangle Y_{L\alpha}(\hat{\vector{r}})\vector{e}_\beta .
\end{eqnarray}
The transition strength of the dipole operator $\hat{M}_D$ for the $0_1^+\rightarrow 1_k^-$ transition
is given as
\begin{eqnarray}
B(D;0_1^+\rightarrow 1_k^-) = |\langle 1_k^-||\hat{M}_D||0_1^+\rangle|^2 .
\end{eqnarray}
The TD and CD operators are IS type rank-1 operators. 
The former can probe the nuclear vorticity as discussed in Ref.~\cite{Nesterenko:2016qiw}. 
The latter is a good probe for the compressional mode and corresponds to the ordinary ISD operator
\begin{eqnarray}
\hat{M}_{\textrm{ISD}}(\mu)=\int d\vector{r}\ \rho(\vector{r})r^3Y_{1\mu}(\hat{\vector{r}})
\end{eqnarray}
with the relation of
\begin{eqnarray}
B(\textrm{CD};0_1^+\rightarrow 1_k^-) &=& \left(\frac{1}{10}\frac{E_k}{\hbar c}\right)^2B(\textrm{ISD};0_1^+\rightarrow 1_k^-), 
\end{eqnarray}
where $E_k$ is the excitation energy of the $1_k^-$.

\section{Effective interaction} \label{sec:effective_int}
The hamiltonian of total system is given as 
\begin{eqnarray}
H = \sum_i t_i - T_G + \sum_{i<j}v_{ij}^{\textrm{coulomb}} + V_{\textrm{eff}} ,
\end{eqnarray} 
where $t_i$ and $T_G$ are the kinetic energy of the $i$th nucleon and the center of mass motion, respectively.
$v_{ij}^{\textrm{coulomb}}$ is the two-body coulomb potential which is approximated by a seven-range Gaussian form.
$V_{\textrm{eff}}$ is effective nuclear interaction of the central and spin-orbit forces.
We use the effective interactions same as Refs.~\cite{Kanada-Enyo:2011ldi,Shikata:2019wdx} for 
$\Beten$, and those as Refs.~\cite{Kanada-Enyo:2017ers,Kanada-Enyo:2019hrm} for $\Oxy$.
That is, for the $\Beten$, we use the Volkov No.2 force~\cite{Volkov:1965zz} with $W = 1-M = 0.6$ and $B=H=0.125$ for the central force and 
the G3RS force~\cite{Yamaguchi:1979hf,Tamagaki:1968zz} with $u_1=-u_2=-1600$\ MeV for the spin-orbit force. 
This parameter set describes well the energy spectra of $\Beten$. 
For the case of $\Oxy$, we use the MV1 central force
with the case1 parametrization of $W = 1-M = 0.62$ and $B=H=0$, and 
the G3RS spin-orbit force  with $u_1=-u_2=-3000$\ MeV, which describe energy spectra of $\Carbon$ with the AMD calculation~\cite{Kanada-Enyo:2006rjf}. 
The Volkov central force is the finite-range 2-body interaction, which describes the $\alpha$-$\alpha$ scattering phase shifts, and often used for light nuclei, but it has an over-binding problem in application to heavy-mass nuclei. 
The MV1 central force consists of the finite-range 2-body and zero-range 3-body terms and can systematically 
reproduce the binding energies of $\alpha$, $\Carbon$, and $\Oxy$~\cite{Kanada-Enyo:2017ers}
without the over-binding problem.

\section{Application to $\Beten$} \label{sec:10Be}

We apply the newly developed method, that is, $\beta$-AMD with K-VAP for $\Beten$. 
We also apply the $\bg$-AMD with P-VAP and K-VAP for comparison, and discuss 
how the new method works for dipole excitations.

\subsection{energy surfaces and Energy spectra}
\begin{figure}[!h]
\begin{center}
\includegraphics[width=12cm]{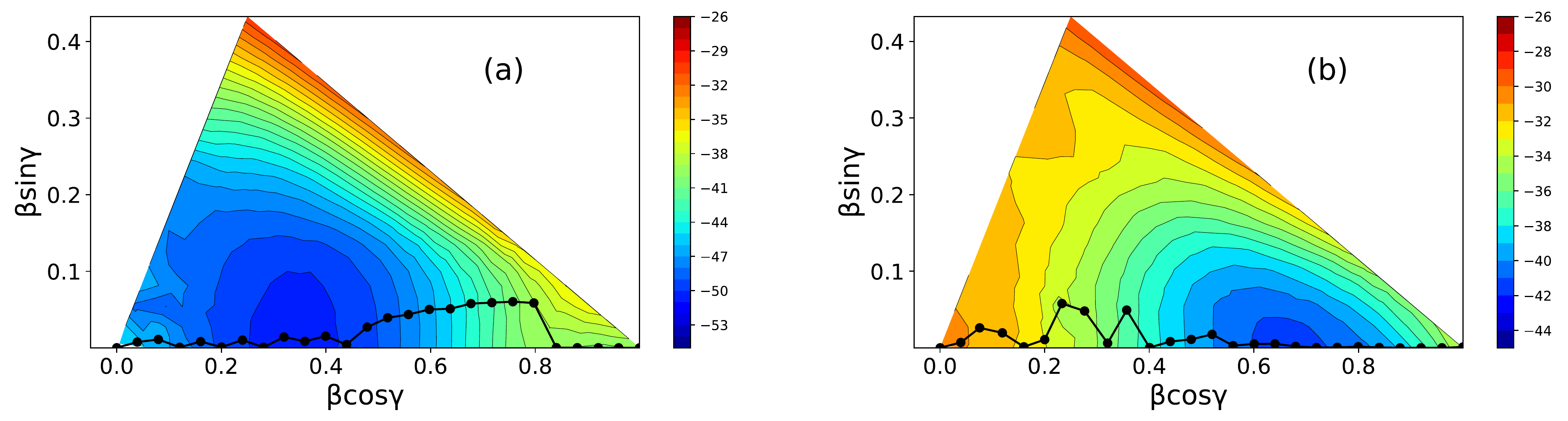}
\includegraphics[width=18cm]{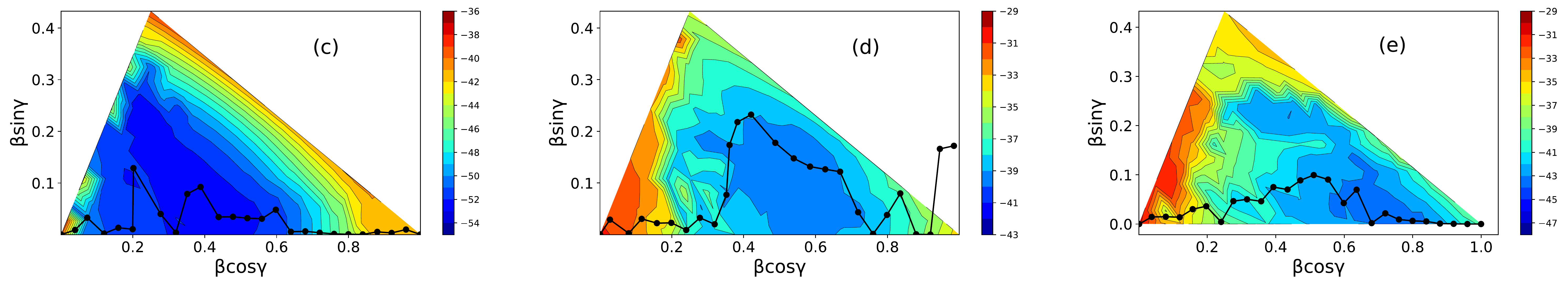}
\end{center}
\caption{(color online) 
energy surfaces of $\Beten$ obtained by $\bg$-AMD with P-VAP and K-VAP.
Panels (a) and (b) are positive- and negative-parity projected energy surfaces, respectively, for the bases of $\bg$-AMD with P-VAP.
Panels (c), (d), and (e) are $K^{\pi}=0^+,\ 0^-,\ $and $1^-$ projected one, respectively, for bases of $\bg$-AMD with K-VAP.
The solid lines with filled circles represent the $(\beta,\gamma)$ paths for $\beta$-AMD bases, which 
are obtained by optimizing $\gamma$ for each $\beta$ value.
Color mapping is shown in the unit of MeV.
}
\label{fig:bg_plane_10Be}
\end{figure}

For detail analysis, we here discuss energy surfaces on the $\beta$-$\gamma$ plane 
obtained by $\bg$-AMD with K-VAP and that with P-VAP.
In Fig.\ \ref{fig:bg_plane_10Be}, the calculated energy surfaces are shown.
Figures\ \ref{fig:bg_plane_10Be}\ (a) and (b) show the P-VAP result for the positive- and negative-parity projections, respectively, and 
Figs.\ \ref{fig:bg_plane_10Be}\ (c), (d), and (e) show the K-VAP result with $K^{\pi}=0^+,\ K^{\pi}=0^-$, and $K^{\pi}=1^-$, respectively.
The K-VAP energy surfaces show a broad flat region around the energy minimum area.
It should be noted that an energy minimum area is located at $\gamma\sim0$ in the  P-VAP energy surfaces 
corresponding to a prolate minimum, but it extends (or shifts) toward a triaxial deformation area ($\gamma\sim 30^{\circ}$) in the $K^{\pi}=0^+$ and $K^{\pi}=0^-$ energy surfaces (Figs.\ \ref{fig:bg_plane_10Be} (c) and (d)). 
This indicates that the axial symmetric deformation is favored before the $K$-projection but the triaxial deformation is favored after the $K$-projection.

The solid lines with filled circle represent the $\gamma$-optimized paths for bases of
$\beta$-AMD, where $\gamma$ is optimized for each $\beta$ value.
In the case of P-VAP, the path goes along the $\gamma=0$ line meaning that $\beta$-AMD with P-VAP gives approximately axial symmetric states.
On the other hand, in the K-VAP result, 
the path goes through a $\gamma\ne 0$ area in the energy valley. 
It means that  triaxially deformed configurations can be obtained by K-VAP even in the $\beta$-AMD. 
As shown later, those triaxial configurations are essential to describe the $K^{\pi}=0^-$ dipole mode.

\begin{figure}[!h]
\begin{center}
\includegraphics[width=12cm]{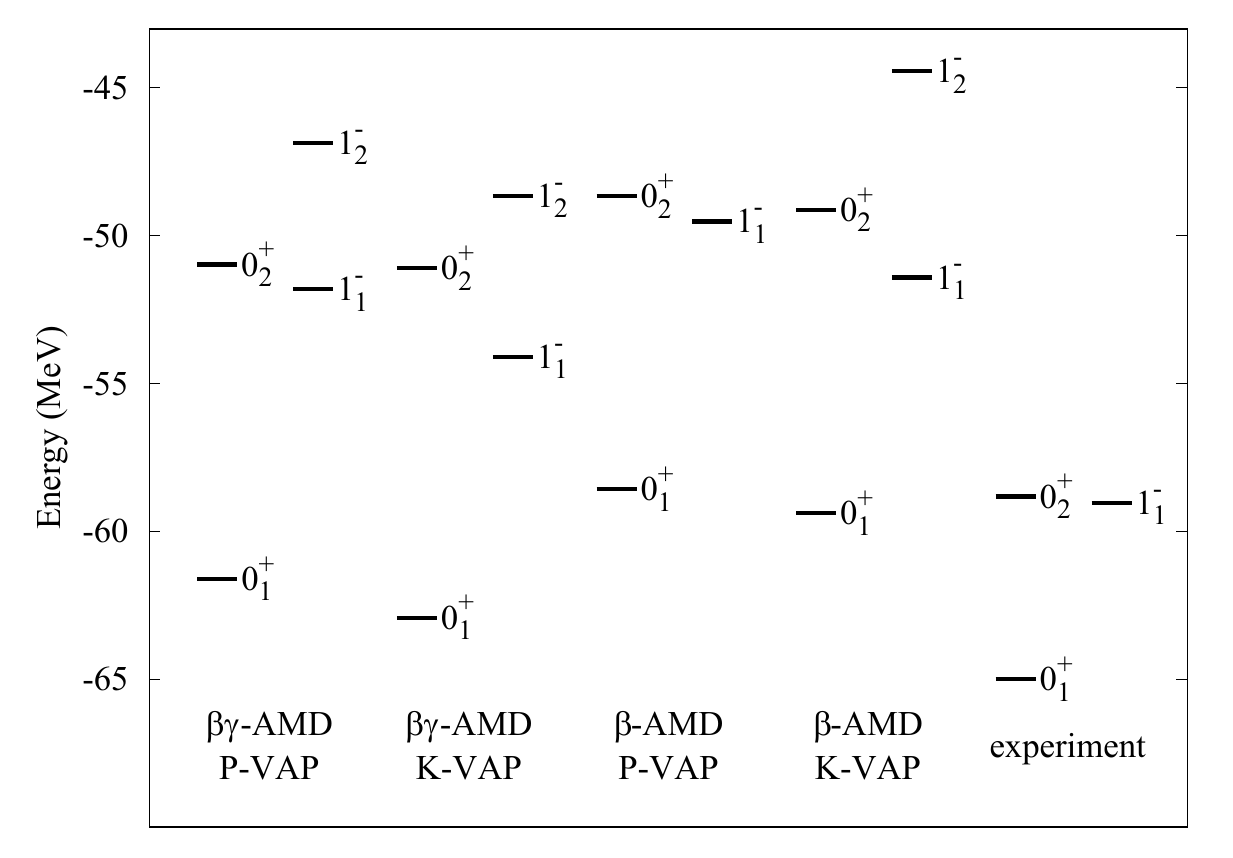}
\end{center}
\caption{
Energy spectra of  $\Beten$ obtained by the GCM calculation of $\bg$-AMD and $\beta$-AMD with 
P-VAP and K-VAP bases. The experimental energy spectra of the corresponding states are shown for 
comparison.
}
\label{fig:spectrum_10Be}
\end{figure}

We perform the GCM calculation for $\beta\gamma$-AMD ($\beta$-AMD) with P-VAP and K-VAP by superposing the obtained basis wave functions to obtain the energy spectra and final wave functions of $\Beten$.
The calculated energy spectra are shown in Fig.\ \ref{fig:spectrum_10Be} compared with the experimental spectra for corresponding states.
In the K-VAP result, energies are globally lower by about a few MeV than the P-VAP result for all states meaning that 
further energy optimization is achieved by the $K$-projection.
For dipole states, two $1^-$ states ($1_{1,2}^-$) are obtained in both the $\bg$-AMD and $\beta$-AMD 
calculations with K-VAP.
However in the case of P-VAP, the $1_2^-$ state is obtained only by $\bg$-AMD but not 
by $\beta$-AMD because the latter calculation ($\beta$-AMD with P-VAP) does not contain triaxially deformed bases 
as shown in the $\gamma$-opimized path in Fig.\ \ref{fig:bg_plane_10Be}\ (b).

\begin{figure}[!h]
\begin{center}
\includegraphics[width=10cm]{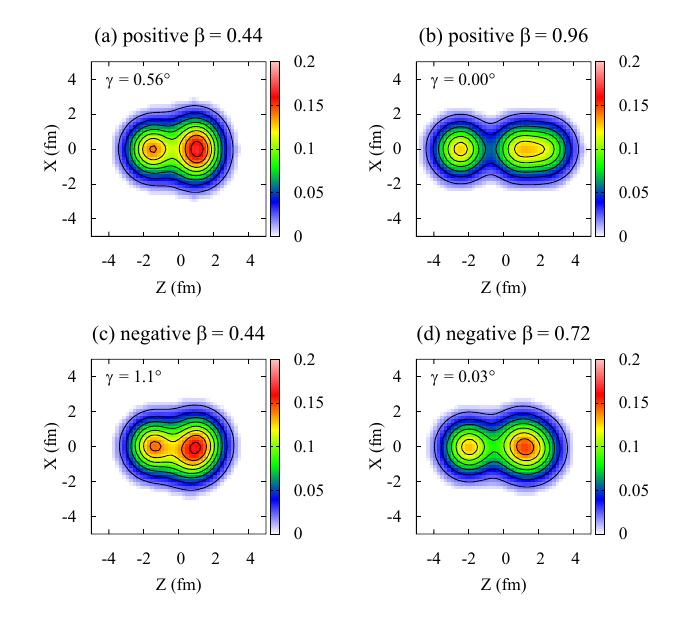}
\end{center}
\caption{(color online) Intrinsic densities of $\Beten$  at given $\beta$ values 
obtained by $\beta$-AMD with P-VAP.
The optimized values of $\gamma$ for each $\beta$ are written in the panels.
Panels (a) and (b) show the densities of the positive-parity bases and (c) and (d) show those of the negative-parity bases.
}
\label{fig:density_P_10Be}
\end{figure}

\begin{figure}[!h]
\begin{center}
\includegraphics[width=10cm]{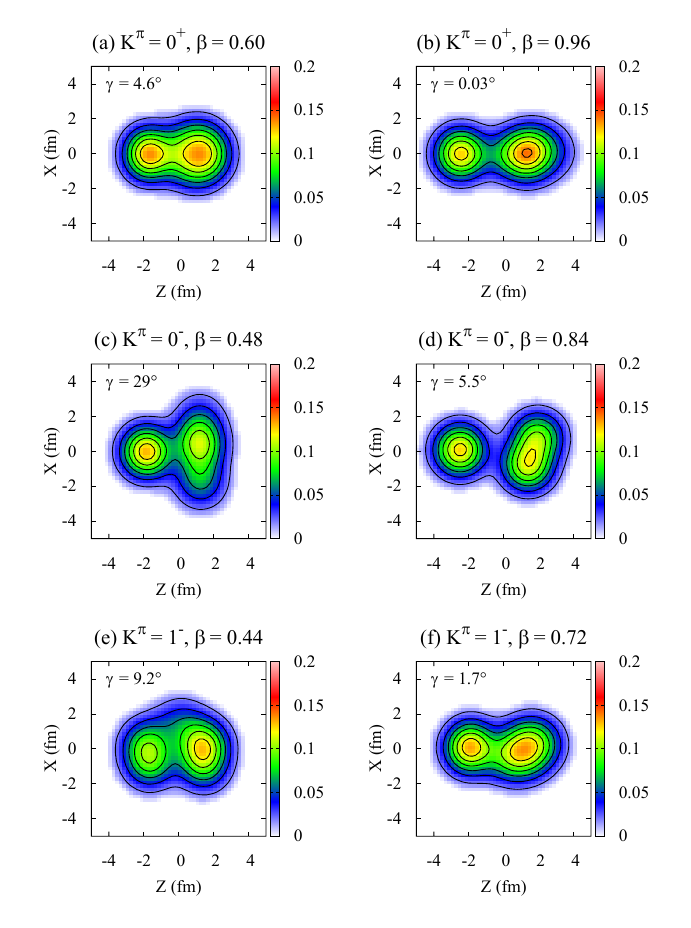}
\end{center}
\caption{(color online) Intrinsic densities of $\Beten$  at given $\beta$ values 
obtained by  $\beta$-AMD with K-VAP.
Top, middle, and bottom panels show the densities of the $K^{\pi}=0^+,\ 0^-,\ $and $1^-$ bases, respectively.
}
\label{fig:density_K_10Be}
\end{figure}

\subsection{properties of dipole excitations}

\begin{table}[htbp]
\begin{center}
\begin{tabular}{llcccc} \toprule
&&$E_x\ $(MeV)&$B(\textrm{TD})\ (\textrm{fm}^4)$&$B(\textrm{CD})\ (\textrm{fm}^4)$&$B(E1)\ (\textrm{fm}^2)$ \\ \midrule
P-VAP&$1_1^-$ &9.05&$2.66\times10^{-3}$&$6.79\times10^{-5}$&$6.42\times10^{-3}$ \\
&$1_2^-$&-&-&-&- \\
&&&&& \\
K-VAP&$1_1^-$ &7.96&$2.39\times10^{-3}$&$2.39\times10^{-4}$&$6.92\times10^{-3}$ \\ 
&$1_2^-$&14.9&$2.59\times10^{-4}$&$2.89\times10^{-4}$&$2.47\times10^{-1}$ \\ 
&&&&& \\
exp.&$1_1^-$ &5.96&-&-&$2.72\times10^{-6}$ \\
&$1_2^-$&-&-&-&- \\
\bottomrule
\end{tabular}
\end{center}
\caption{The excitation energies and transition strengths in $\Beten$ obtained by $\beta$-AMD with P-VAP and K-VAP, and experimental data~\cite{PhysRevC.80.034318}.
}
\label{tab:strength_10Be}
\end{table} 

In this subsection, we discuss properties of dipole excitations mainly based on the result of $\beta$-AMD with K-VAP.
The excitation energies and dipole transition strengths for the $E1$, CD, and TD operators are shown in Table\ \ref{tab:strength_10Be}.
The intrinsic densities of basis wave functions obtained by $\beta$-AMD with K-VAP and P-VAP 
are shown in  Fig.\ \ref{fig:density_K_10Be} and Fig.\ \ref{fig:density_P_10Be}, respectively.
In the K-VAP result, the $0_1^+$ state has a largest overlap with the configuration at $(\beta,\gamma)=(0.60,4.6^{\circ})$ in Fig.\ \ref{fig:density_K_10Be}\ (a), whereas the $1_1^-$ state dominantly contains the $K=1$ component of the intrinsic state at $(\beta,\gamma)=(0.72,1.7^{\circ})$ in Fig.\ \ref{fig:density_K_10Be}\ (f). 
They show approximately axial-symmetric shapes ($\gamma\sim 0$) with an $\alpha+\Hesix$-like structure. 
In these states, two clusters are close to each other and 
not well-developed. 
The $1_1^-$ state has the strong TD transition and can be regarded as the $K=1$ excitation built on the compact $\alpha+\Hesix$ state. This remarkable TD strength is caused by vortical current in the transition current density and it is considered that the excitation mode to the $1_1^-$ is the vortical mode consistent with that discussed in Refs.~\cite{Kanada-Enyo:2017uzz,Shikata:2019wdx}.
For  the $0_1^+$ and $1_1^-$ states, the P-VAP result shows similar features 
to the K-VAP result in the TD transition, but differs from the K-VAP result in the CD strength.
Namely, the K-VAP calculation shows three times larger CD strength for the $1^-_1$ than the P-VAP calculation 
because of significant mixing of the $K=0$ component of triaxially deformed configurations 
(see Fig.\ \ref{fig:density_K_10Be}\ (c)). 

The $1_2^-$ state is dominantly formed by the $K=0$ component of  
triaxially deformed configurations obtained by $\beta$-AMD with K-VAP for the $K^\pi=0^-$. It 
has the dominant overlap  ($\sim$ 70\ \%) with the basis wave function at $(\beta,\gamma)=(0.84,5.5^{\circ})$
with a remarkably developed $\alpha+\Hesix$ cluster structure
as shown in Fig.\ \ref{fig:density_K_10Be}\ (d). 
It also has significant overlap ($\sim$ 52\ \%) with the base at $(\beta,\gamma)=(0.48,29^{\circ})$ 
in Fig.\ \ref{fig:density_K_10Be}\ (c). In these configurations, 
the $\Hesix$ cluster is deformed in a tilted or perpendicular orientation from the $Z$ axis
contributing the triaxiality of the total system. The $K=0$ component projected from these triaxial configurations play a crucial role in generating the $1_2^-$ state. 

In the transition properties,  the remarkable $E1$ strength is obtained for the $1_2^-$:
Energy-weighted strength ($EB(E1:0_1^+\rightarrow 1_2^-)$) accounts for about 10\ $\%$ of the TRK sum rule.
It also has the significant CD transition strength because of the $K=0$ component of the developed $\alpha+\Hesix$ cluster structure. 
Furthermore, as a result of mixing of the large triaxial configuration in Fig.\ \ref{fig:density_K_10Be}\ (c), the CD strength and also  $E1$ strength are 
 enhanced because the excitation from the ground state to such triaxial configuration involves spatial expansion of excess neutron distribution in $\Hesix$, which induces the $E1$ and CD strengths as described in Ref.~\cite{Shikata:2019wdx}.

In the application to $\Beten$, we can say that $\beta$-AMD combined with K-VAP is useful to efficiently 
generate important bases for the low-energy dipole excitations, $1^-_1$ and $1^-_2$ states, and can obtain the equivalent result
to the  $\beta\gamma$-AMD. In particular, the K-VAP treatment can get 
the triaxial configurations within the $\beta$-AMD, which are essential for the $1^-_2$ state ($K=0$ mode).
This is an advantage against $\bg$-AMD to save computational costs, and superior to the standard $\beta$-AMD with the P-VAP framework, 
which favors axial-symmetric configurations rather than axial-asymmetric ones. 

\section{Results of $\Oxy$} \label{sec:16O}
In the previous section, we demonstrated applicability of $\beta$-AMD with K-VAP in description of 
dipole excitations in $\Beten$.
We apply the same method to $\Oxy$ and present the obtained results for LED excitations
in this section.
\subsection{energy surfaces and intrinsic structures}

\begin{figure}[!h]
\begin{center}
\includegraphics[width=\hsize]{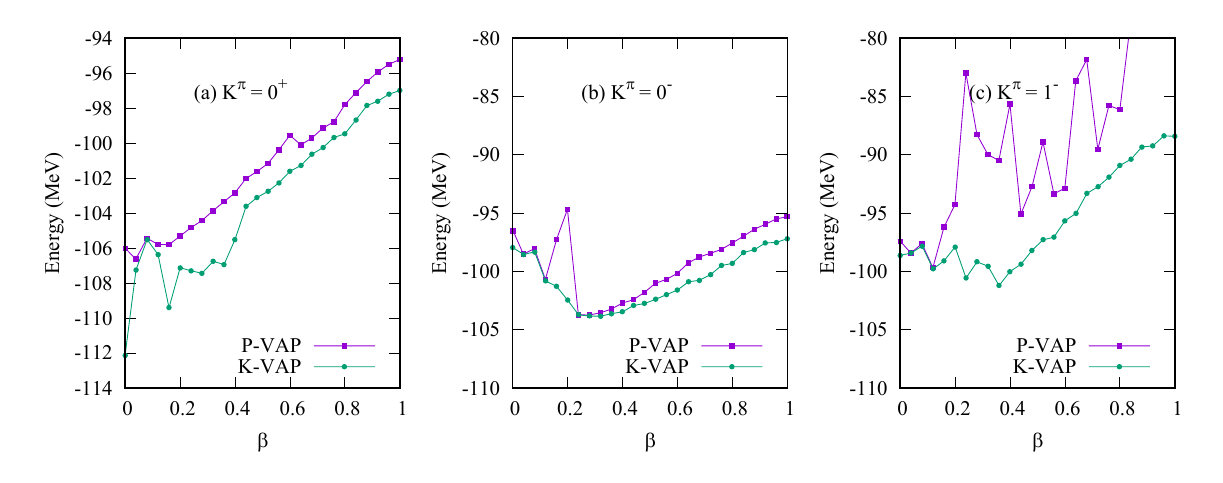}
\end{center}
\caption{(color online) The energy surfaces obtained by $\beta$-AMD as a function of $\beta$ in $\Oxy$.
Panels (a), (b), and (c) show the $K^{\pi}=0^+,\ 0^-,$ and $1^-$ projected energy surfaces, respectively.
The solid lines with filled circle are the result of K-VAP, and those with filled square represent the P-VAP result.
}
\label{fig:ES_bc_16O}
\end{figure}

$K$-projected energy curves obtained by $\beta$-AMD with K-VAP are shown in Fig.\ \ref{fig:ES_bc_16O}
compared with the P-VAP result. 
Figures\ \ref{fig:ES_bc_16O}\ (a), (b), and (c) show the $K^{\pi}=0^+,\ 0^-,\ $and $1^-$ projected energy surfaces, respectively.
In all the cases, the energy minimum corresponds to the normal deformation(ND) around $\beta=0.2$--$0.4$ indicating
a relatively smaller deformation in $\Oxy$ than $\Beten$.
In the negative-parity results, 
the energy difference between the $K=0$ and $K=1$ bases is only a few MeV at the energy minimum, 
but it gets larger as $\beta$ increases: the energy curve for $K^{\pi}=0^-$($K^{\pi}=1^-$) bases 
is soft(steep) against $\beta$.

\begin{figure}[!h]
\begin{center}
\includegraphics[width=\hsize]{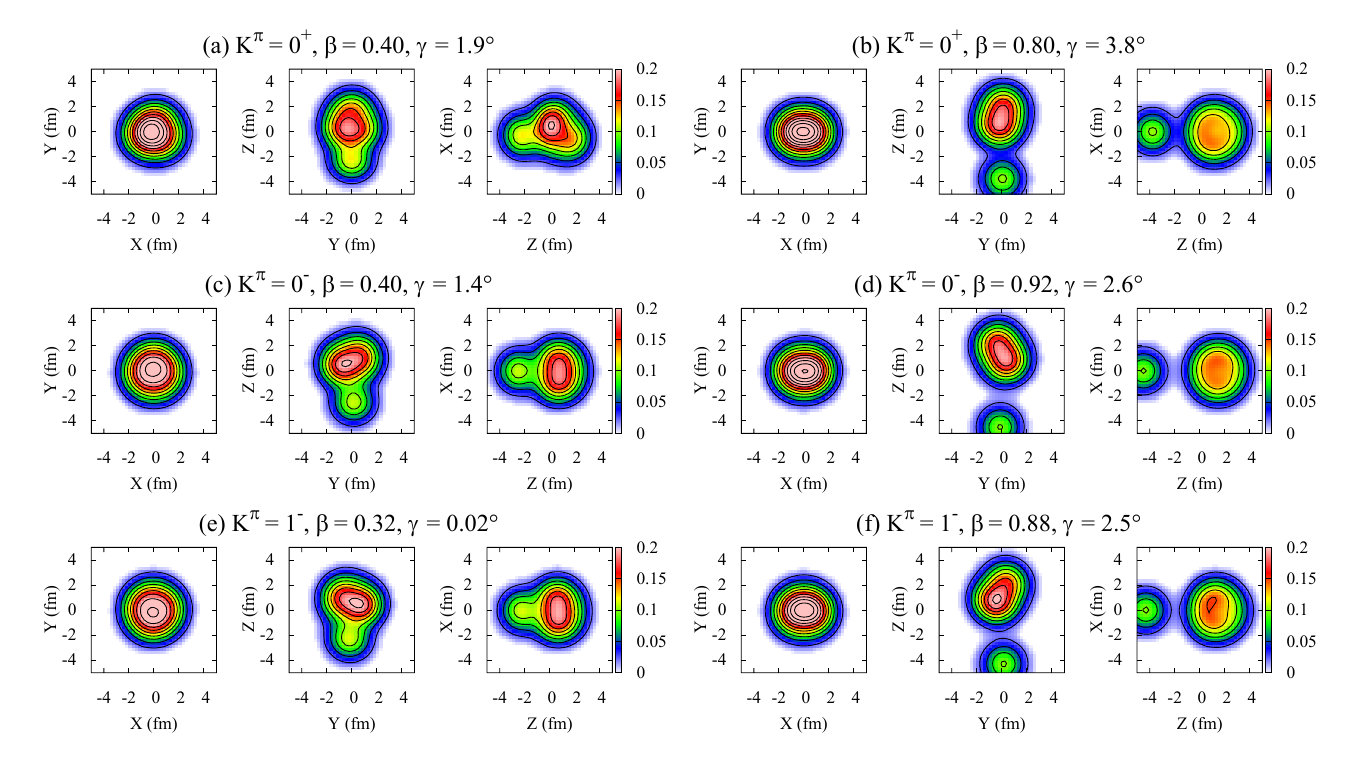}
\end{center}
\caption{(color online) Intrinsic densities of  $\Oxy$ at given $\beta$ values 
obtained by the $\beta$-AMD with K-VAP.
Top, middle, and bottom panels show the densities of the $K^{\pi}=0^+,\ 0^-,\ $and $1^-$ bases, respectively.
}
\label{fig:density_K_16O}
\end{figure}

In Fig. \ref{fig:density_K_16O}, intrinsic densities of K-VAP bases are shown.
In all cases of $K^{\pi}=0^+,$ $0^-$, and  $1^-$, an $\alpha$ cluster is formed at the surface of 
a $^{12}$C core in the ND region, and it develops as $\beta$ increases.  
The $K^{\pi}=0^+$ bases around the energy minimum at $\beta\sim0.4$ has a tetrahedral-like structure
as shown in Fig.\ \ref{fig:density_K_16O}\ (a), which contributes to the ground state obtained by the 
GCM calculation. 
In a large $\beta$ region, the $K^{\pi}=0^+$ bases show a developed $\Carbon+\alpha$ cluster structure (see Fig.\ \ref{fig:density_K_16O}\ (b)), 
which dominantly contributes to the $0_2^+$ state.
In the $K^{\pi}=0^-$ and $K^{\pi}=1^-$ bases, 
intrinsic configurations at $\beta\sim 0.4$ (ND region) are similar to each other as shown  in 
Figs.\ \ref{fig:density_K_16O} (c) and (e).
These two bases have large overlaps with the $1_1^-$ state obtained by the GCM calculation.
At large $\beta$ values, $K^{\pi}=0^-$ and $K^{\pi}=1^-$ bases have 
developed $\Carbon+\alpha$ cluster configurations, where 
the $\Carbon$ cluster is deformed with tilted orientation against the $\Carbon$-$\alpha$ $(Z)$ axis
as shown  in 
Figs.\ \ref{fig:density_K_16O} (d) and (f). 
Consequently, the total system becomes axial asymmetric. 
The $K^{\pi}=0^-$ bases with this axial asymmetric cluster structure 
play an important role to construct the $1_2^-$ state as discussed later. 

\begin{figure}[!h]
\begin{center}
\includegraphics[width=\hsize]{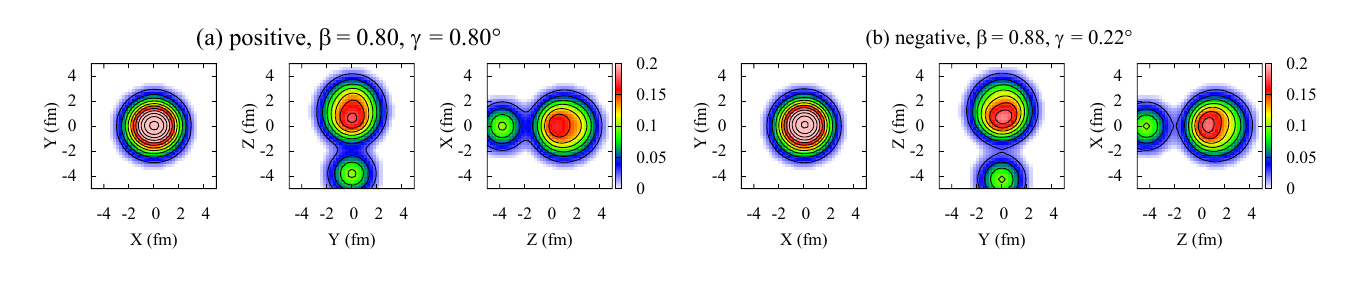}
\end{center}
\caption{(color online) Intrinsic densities of  $\Oxy$ at given $\beta$ values obtained by $\beta$-AMD with P-VAP.
The optimized values of $\gamma$ for each $\beta$ are also written.
Panels (a) and (b) show the densities of the positive- and negative-parity bases, respectively.
}
\label{fig:density_P_16O}
\end{figure}

\subsection{GCM results and LED strengths}

\begin{figure}[!h]
\begin{center}
\includegraphics[width=12cm]{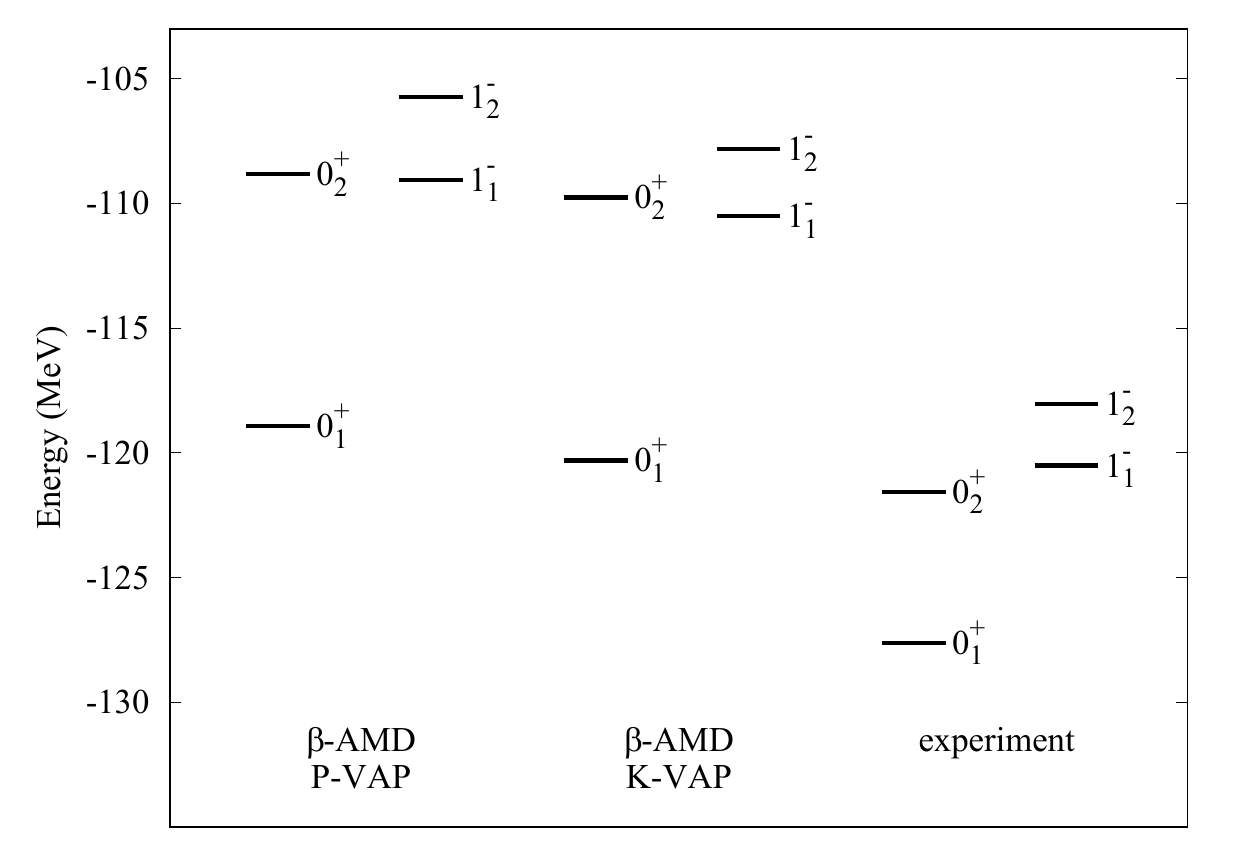}
\end{center}
\caption{Energy spectra of  $\Oxy$ obtained by the GCM calculation of $\beta$-AMD with 
P-VAP and K-VAP bases. The experimental energy spectra of the corresponding states are shown for 
comparison.
}
\label{fig:spectrum_16O}
\end{figure}

\begin{figure}[!h]
\begin{center}
\includegraphics[width=\hsize]{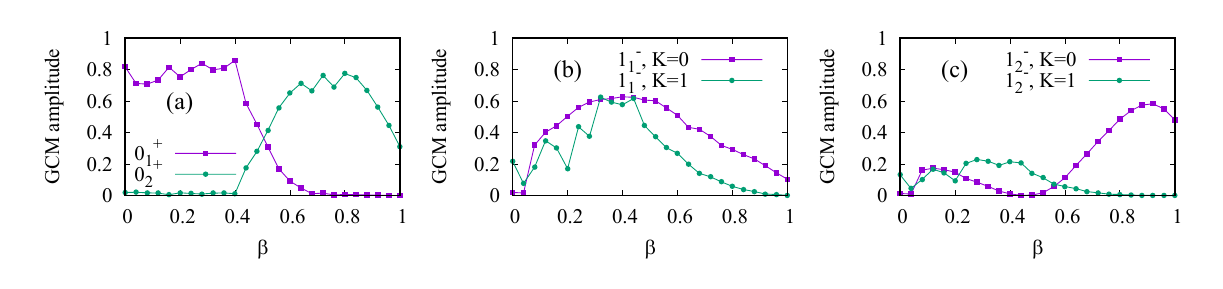}
\end{center}
\caption{ (color online)
GCM amplitudes as functions of $\beta$ for the K-VAP results of $\Oxy$.
In panel (a), GCM amplitudes for the $0_1^+$ and $0_2^+$ states are plotted by square 
and circle points, respectively.
Panels (b) and (c) show the amplitude for the $1_1^-$ and $1_2^-$, respectively. 
The $K=0$ component of GCM amplitude for the $K^{\pi}=0^-$ basis and  
the $K=1$ component of GCM amplitude for the $K^{\pi}=1^-$ basis are plotted by 
square and circle points, respectively.
}
\label{fig:SO_16O}
\end{figure}

Energy spectra obtained by the GCM calculation with K-VAP bases are shown in Fig.\ \ref{fig:spectrum_16O}
compared with the experimental data.
In Fig.\ \ref{fig:SO_16O}, GCM amplitudes, which are given by squared overlaps of the  GCM states with 
K-VAP bases, are shown as functions of $\beta$.
In Fig.\ \ref{fig:SO_16O}\ (a),
the GCM amplitudes for the $0_1^+$ and $0_2^+$ states are shown by square and circle points, respectively.
In Figs.\ \ref{fig:SO_16O}\ (b) and (c), the GCM amplitudes for the $1_1^-$ and $1_2^-$ states are shown, respectively, with square(circle) points for $K^{\pi}=0^-(1^-)$ bases.

The calculated binding energy of the ground state is $120.3$\ MeV which slightly underestimates 
the experimental value $127.6$\ MeV.
The ground state has the largest GCM amplitude as $\sim 90$\ $\%$
with the base at $\beta=0.4$ as shown in Fig.\ \ref{fig:density_K_16O}\ (a).
The lowest $1^-$ state at the excitation energy 9.8\ MeV largely contains the $K^\pi=0^-$ and $K^\pi=1^-$ components in the $\beta\sim0.4$ (ND) region (see Fig.\ \ref{fig:SO_16O}\ (b)).
The $0_2^+$ state at 10.55\ MeV and the $1_2^-$ state at 12.47\ MeV 
are constructed from the $K^{\pi}=0^+$ and  $K^{\pi}=0^-$ bases with the developed
$\alpha + \Carbon$ cluster structure, respectively, 
as can be seen in their large 
GCM amplitudes at 0.6$\lesssim\beta\lesssim0.8$ region (see Figs.\ \ref{fig:SO_16O}\ (a) and (c)). 
This result is consistent with the cluster model assignment of the 
 $0_2^+$ and  $1_2^-$  states as the band-head states of the parity doublet
($K^{\pi}=0^+$ and $K^{\pi}=0^-$) $\alpha + \Carbon$ cluster bands.

\begin{table}[htbp]
\begin{center}
\begin{tabular}{llccc} \toprule
&&$E_x\ $(MeV)&$B(\textrm{TD})\ (\textrm{fm}^4)$&$B(\textrm{CD})\ (\textrm{fm}^4)$ \\ \midrule
P-VAP&$1_1^-$ &9.88&$2.93\times10^{-3}$&$3.49\times10^{-3}$ \\
&$1_2^-$&13.20&$3.98\times10^{-3}$&$1.65\times10^{-5}$ \\
&&&& \\
K-VAP&$1_1^-$ &9.78&$3.62\times10^{-3}$&$2.87\times10^{-3}$ \\
&$1_2^-$&12.47&$3.80\times10^{-3}$&$9.35\times10^{-5}$ \\ \bottomrule
\end{tabular}
\end{center}
\caption{The excitation energies and transition strengths in $\Oxy$.}
\label{tab:strength_16O}
\end{table} 

The calculated dipole transition strengths are shown in Table\ \ref{tab:strength_16O}.
The present calculation obtains significant TD and CD strengths of the $1_1^-$.
This result for the $1_1^-$ state is qualitatively 
consistent with the previous AMD result obtained by the variation after spin-parity projections
\cite{Kanada-Enyo:2017ers}. The TD strength is also shared by the $1_2^-$ state
via mixing of the TD mode and $\alpha + \Carbon$ cluster mode.
More details of the two dipole modes are discussed later. 

\subsection{comparison between K-VAP and P-VAP methods}
For comparison, we also perform $\beta$-AMD with P-VAP of $\Oxy$ for comparison with the 
present K-VAP method.
The $K$-projected energy curves obtained from P-VAP bases are shown in Fig.\ \ref{fig:ES_bc_16O}
by square points, and intrinsic densities of P-VAP bases are shown in Fig.~\ref{fig:density_P_16O}. 
The K-VAP energy curves are generally lower than the P-VAP result
indicating that additional energy gains are obtained with the K-VAP method.
In particular, the significant energy gain is obtained for $K^{\pi}=1^-$ bases because P-VAP
bases for the negative-parity projection contains the dominant  $K=0$ component but minor 
$K=1$ component. This is a general trend of the P-VAP method, which favor
axial-symmetric configurations, i.e.,  $K=0$ dominant bases in various nuclei. 
Indeed, as shown in the intrinsic densities of Fig. \ref{fig:density_K_16O}, P-VAP
bases show approximately axial symmetric configurations.

In the GCM energies of  $\Oxy$ states obtained by superposing K-VAP or P-VAP bases, 
all the GCM states obtained by K-VAP bases gain $1$-$2$ MeV energies  compared with the P-VAP result. 
For example, the binding energy of the  K-VAP result is 1.4 MeV larger than the P-VAP result.
Further energy gain is obtained for $\Carbon+\alpha$ ($K^{\pi}=0^-$) cluster band because of the rotational degree of freedom of the deformed $\Carbon$ cluster in K-VAP bases.
In Table\ \ref{tab:strength_16O}, the properties of dipole excitations are compared between the 
P-VAP and K-VAP results. The two methods give qualitatively similar results for excitation energies and 
TD and CD transition strengths of the $1^-_1$ and $1^-_2$ states. It means that, 
in the case of $\Oxy$, the essential bases for low-energy dipole states can be obtained 
by the P-VAP method, even though the K-VAP method obtains some amount of extra energy gain due to higher order effects.

\subsection{excitation modes and origin of strengths}
We analyze dipole modes and discuss origins of the significant dipole strengths in the present K-VAP result 
of $\Oxy$. For this aim, 
we define an ``intrinsic state'' of each GCM state as a specific base state having the largest GCM amplitude.
For the $0_1^+$ state, the intrinsic state is defined as the $K^{\pi}=0^+$ base at $\beta = 0.4$ shown in Fig.\ \ref{fig:density_K_16O}\ (a).
In the $1_1^-$ state, the $K^{\pi} = 0^-$ and $K^{\pi} = 1^-$ bases at the $\beta\sim 0.4$ (ND) region are strongly mixed and  
both have large GCM amplitudes, and therefore, 
two choices of the intrinsic states are possible: one is the  $K^{\pi} = 0^-$ base 
at $\beta=0.40$ shown in Fig.\ \ref{fig:density_K_16O}\ (c) 
and the other is the $K^{\pi} = 1^-$ base at $\beta=0.32$ shown in Fig.\ \ref{fig:density_K_16O}\ (e).
We call these intrinsic states as $\IntGS$, $\IntNDzero$, and $\IntNDone$.
The $1_2^-$ contains mainly the $K^{\pi}=0^-$ components with the developed 
$\alpha+\Carbon$ structure. 
Therefore, as the intrinsic state of the $1_2^-$ state, 
we choose the  $K^{\pi}=0^-$  base at $\beta= 0.92$ (Fig.\ \ref{fig:density_K_16O}\ (f)), which we label as $\IntCL$.

In order to clarify the dipole excitation modes, we calculate transition current densities $\vector{j}^K(\vector{r})$ 
and local matrix elements of the CD and TD operators $\hana{M}_D^K(\vector{r})\ (D = \textrm{CD}\ \textrm{or}\ \textrm{TD})$
for transitions between $K$-projected states in the intrinsic frame defined as, 
\begin{eqnarray}
\vector{j}^K(\vector{r}) &\equiv & \langle f^K|\hat{\vector{j}}_{\textrm{nucl}}(\vector{r})|i\rangle ,\\
\hana{M}_{\textrm{TD}}^{K=0}(\vector{r})&=& \frac{1}{c}\left[ (2X^2+2Y^2+Z^2)j^{K=0}_Z - ZXj^{K=0}_X - YZj^{K=0}_Y \right],\label{eq:SD_TD_0} \\ 
\hana{M}_{\textrm{TD}}^{K=1}(\vector{r})&=& \frac{1}{c}\left[ (X^2+2Y^2+2Z^2)j^{K=1}_X - XY j^{K=1}_Y - ZXj^{K=1}_Z \right],\label{eq:SD_TD_1} \\
\hana{M}_{\textrm{CD}}^{K=0}(\vector{r})&=& \frac{1}{c}\left[ -(X^2+Y^2+3Z^2)j^{K=0}_Z - 2ZXj^{K=0}_X - YZj^{K=0}_Y \right],\label{eq:SD_CD_0} \\
\hana{M}_{\textrm{CD}}^{K=1}(\vector{r})&=& \frac{1}{c}\left[ -(3X^2+Y^2+Z^2)j^{K=1}_X - 2XY j^{K=1}_Y - 2ZXj^{K=1}_Z \right],\label{eq:SD_CD_1}
\end{eqnarray}
where the initial state $|i\rangle$ is taken to be $\hat{P}^{K=0}|\IntGS\rangle$, and the final state $|f^K\rangle$ is chosen to be either of $\hat{P}^{K}|f\rangle$ with 
$|f\rangle=|\IntNDzero\rangle$, $|\IntNDone\rangle$, and $|\IntCL\rangle$.
$\hana{M}_{\textrm{TD(CD)}}^{K}$ is the local matrix element at $\vector{r}$ corresponding to integrand of the TD(CD) strength
and called "TD(CD) strength density" in the present paper. 
Here, we specify the $\mu$ component of the dipole 
operator in the TD strength, $\vector{Y}_{11\mu=0}$ with the label $K=0$
and $-(\vector{Y}_{11\mu=1}-\vector{Y}_{11\mu=-1})/\sqrt{2}$ with $K=1$, 
and that in the CD strength, $Y_{1\mu=0}$ with the label $K=0$ and $-(Y_{1\mu=1}-Y_{1\mu=-1})/\sqrt{2}$ with $K=1$.

\begin{figure}[!h]
\begin{center}
\includegraphics[width=13cm]{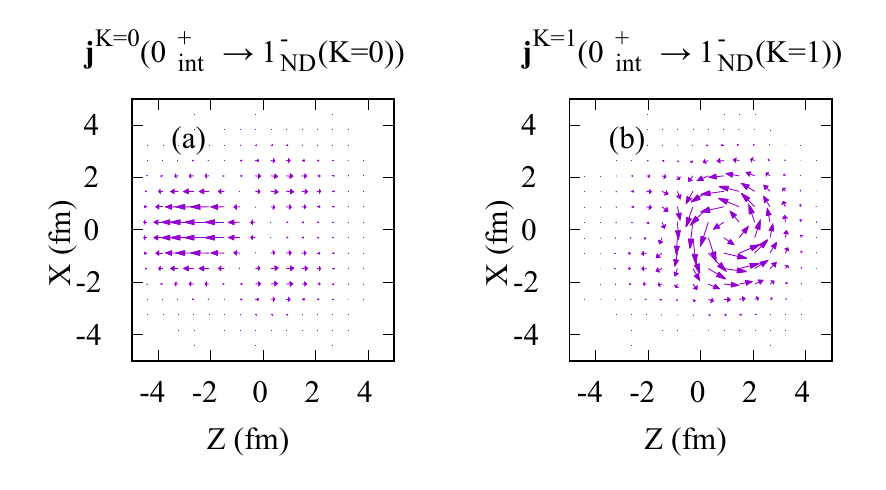}
\end{center}
\caption{(color online) The transition current densities for the $K$-projected intrinsic states of $\Oxy$.
Panels (a) and (b) show the transition currents to the $\IntNDzero$ and $\IntNDone$, respectively.
The densities are plotted on the $Z$-$X$ plane.
}
\label{fig:current_16O}
\end{figure}

The calculated transition current densities for  $\IntGS\to \IntNDzero$ and $\IntGS\to \IntNDone$ transitions
are shown in Fig.\ \ref{fig:current_16O}.
In the vector plot of $\vector{j}^K(\vector{r})$ in Fig.\ \ref{fig:current_16O}, 
one can see two kinds of dipole modes characterized by $K$ quanta.
One is the $K=0$ oscillation mode and the other is the $K=1$ vortical mode.
The former shows translational current along the $Z$ direction (see Fig.\ \ref{fig:current_16O}\ (a)) caused by the $\alpha$ cluster oscillation relative to the $^{12}$C$(3\alpha$) core, which is understood as a $K^\pi=0^-$ excitation of the relative motion between two asymmetric clusters.
The latter shows the remarkable vortical current (see Fig.\ \ref{fig:current_16O}\ (b)) arising from the tilted $\Carbon$ configuration in Fig.\ \ref{fig:density_K_16O}\ (e).
It should be noted that, in the $J^\pi=1^-$ projected states, 
these two dipole modes of the $K=0$ and $K=1$ components are not orthogonal to but somewhat overlap
with each other, 
and both are significantly contained in the $1_1^-$. 
It means that the $1_1^-$ has the dual nature of dipole excitations characterized by $K$-quanta:
the $K=1$ vortical mode and the $K=0$ oscillation mode.

\begin{figure}[!h]
\begin{center}
\includegraphics[width=\hsize]{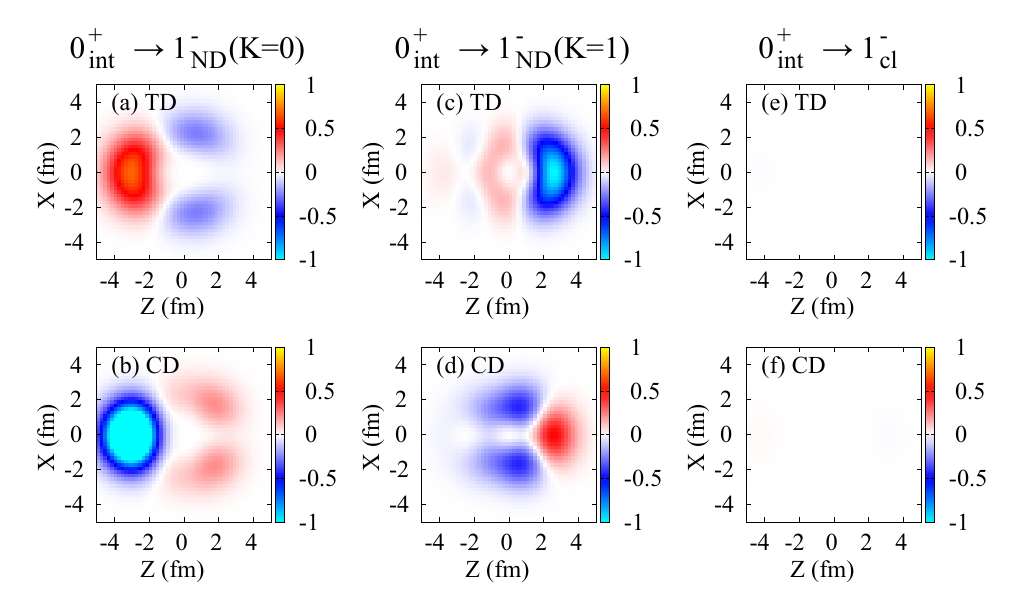}
\end{center}
\caption{(color online) The heat maps for the TD and CD strength densities in $\Oxy$ with respect to the $K$-projected intrinsic states in $Z$-$X$ plane at $Y=0$.
Panels (a)-(b) and (c)-(d) show the matrix elements from the $\IntGS$ to the $\IntNDzero$ and $\IntNDone$, respectively.
Panels (e) and (f) show the one to the $\IntCL$.
Color mapping is shown in the range $[-1\ \textrm{fm},1\ \textrm{fm}]$.
}
\label{fig:strength_density_16O}
\end{figure}

In order to discuss contributions of these two dipole modes to the TD and CD transition strengths, 
we show in Fig.\ \ref{fig:strength_density_16O} the strength densities in the $Z$-$X$ plane at $Y=0$, $\hana{M}_{\textrm{TD}}^{K}(X,0,Z)$ and $\hana{M}_\textrm{CD}^{K}(X,0,Z)$, defined in eqs.\ \eqref{eq:SD_TD_0}-\eqref{eq:SD_CD_1}.
Upper and lower panels of Fig.\ \ref{fig:strength_density_16O} show 
the calculated results of the TD and CD strengths, respectively,
for the transitions from the $\IntGS$ to the three $1^-$ states: $\IntNDzero$, $\IntNDone$, and $\IntCL$.
The $K=0$ oscillation, i.e., the $\alpha$ cluster oscillation, generates the strong CD strength density as seen in Fig.\ \ref{fig:strength_density_16O} (b), 
whereas the $K=1$ vortical current contributes to the TD strength density as can be seen in Fig.\ \ref{fig:strength_density_16O} (c). 
Compared with the $\IntGS\to \IntNDzero$ and $\IntGS\to \IntNDone$ transitions, 
the strength densities $\hana{M}_D^K(\vector{r})$ are very weak in the $\IntCL$ excitation (see Figs.\ \ref{fig:strength_density_16O}\ (e) and (f)) because of the small overlap between the initial and final states, that is, the compact ground state and the developed cluster state.

In the present analysis of the dominant intrinsic configurations,  
the pure $K^\pi=0^-$ cluster excitation does not generate strong TD and CD strengths. 
The $K^\pi=0^-$ cluster component is the main component of the $1_2^-$ state. 
Nevertheless, the $1_2^-$ state has the significant TD strength in the GCM result as shown previously, because it is not a pure $K^\pi=0^-$ cluster state
but contains significant mixing ($\sim 30\%$) of the $K=1$ component at the $\beta\sim 0.4$ (ND) region, which contributes to the 
significant TD strength.
In other words, the TD strength of the $1_2^-$ state comes from mixing of the $K^\pi=1^-$ vortical mode in the dominant 
$K^\pi=0^-$ cluster mode.

The TD strength of the  $1_2^-$ state is sensitive to the mixing ratio of the $K=1$ configuration.
To see the sensitivity, we artificially vary the mixing ratio by changing the strength $V_{LS}$ of the spin-orbit force 
as $V'_{LS} = \lambda V_{LS}$. We perform the GCM calculation with 
$\lambda=0.8$ and 0.9 for 80\% and 90\% weaker spin-orbit strengths. As $\lambda$ decreases, 
the energy interval between the $1_1^-$ and $1_2^-$ states increases and the mixing becomes small. 
As a result, the TD strength is concentrated on the $1_1^-$ state. 
For instance, in the 
$\lambda = 0.8$ case, the TD strengths are
 $B(\textrm{TD};0_1^+\rightarrow 1_1^-)=8.30\times10^{-3}$\ $\textrm{fm}^4$ and $B(\textrm{TD};0_1^+\rightarrow 1_2^-)=1.41\times10^{-3}$\ $\textrm{fm}^4$.
From this analysis, one can see that the present result 
of $B(\textrm{TD};0_1^+\rightarrow 1_1^-)=3.62\times10^{-3}$\ $\textrm{fm}^4$ and $B(\textrm{TD};0_1^+\rightarrow 1_2^-)=3.80\times10^{-3}$\ $\textrm{fm}^4$
for the original strength ($\lambda = 1.0$) is a transient situation of the 
level crossing between the lower $K^\pi=1^-$ vortical mode and higher $K^\pi=0^-$ cluster mode, 
where the TD strength from the $K^\pi=1^-$ vortical mode is fragmented into the $1^-_1$ and $1^-_2$ states. 
Since there remains model ambiguity in effective interaction parameters in the present framework, 
we can not give a conclusion for the mixing ratio.
In order to understand detailed properties of low-energy dipole excitations in $\Oxy$, 
further experimental informations are needed.

\section{Summary}\label{sec:summary}

We proposed a new method, $\beta$-AMD with K-VAP, in which variation is performed after the $K$-projection as well as the parity projection under the $\beta$-constraint, and applied this method for low-energy dipole excitations of $\Beten$ and $\Oxy$.

In the application to $\Beten$, we showed that the new method works well to efficiently obtain essential 
configurations for low-energy dipole excitations and can obtain 
better results compared with the standard $\beta$-AMD with P-VAP (without 
$K$-projection). In particular, 
$\beta$-AMD with K-VAP can describe triaxially deformed configurations, 
whereas the $\beta$-AMD with P-VAP favors only the axial-symmetric ($\gamma\sim0$) configurations.
In the GCM calculation, the present $\beta$-AMD with K-VAP can obtain  
the equivalent result to the GCM result of full  $\bg$-AMD configurations
for the $1_1^-$ and $1_2^-$ states.
It indicates that the present method of $\beta$-AMD with K-VAP is a useful approach to save numerical cost
in describing low-energy dipole excitations.

In the application of $\Oxy$, the present method obtains two LED states, $1_1^-$ and $1_2^-$.
The ground state and $1_1^-$ state are dominantly contributed by small deformation bases and 
the $1_1^-$ state possesses the significant TD and CD transition strengths.
On the other hand, the $1_2^-$ state is mainly composed by $\alpha + \Carbon$ cluster developed bases and regarded as the band-head state of the $K^\pi=0^-$ cluster band. 
By analysis of the strengths and current densities of these two types (TD and CD) of dipole transitions,  
we discussed properties of the dipole excitations and showed that the $K=0$ and $K=1$ natures play important roles. 
Namely, the $1_1^-$ state has the duality of the $K=0$ and $K=1$ modes: the $K=0$ oscillation mode is described by relative motion between asymmetric clusters, and $K=1$ vortical mode is generated by the tilted configuration of the deformed $\Carbon$ cluster.
The $1_2^-$ state is dominantly described by the $K=0$ dipole motion of the $\alpha$ cluster relative to the $\Carbon$ cluster.

The present method was found to be a useful tool for study of LED in light nuclei. 
We will apply this method to $Z\neq N$ nuclei in $sd$-shell region such as $^{18}$O and $^{20}$O in future study.
It is also an remaining issue to discuss experimental signal of the TD mode. 
For instance, transverse charge form factors would be useful experimental information as a
sensitive probe for the TD transition strengths as pointed out by Nesterenko {\it et al.} \cite{Nesterenko:2019dnt}.

\section*{Acknowledgement}

The authors thank to Dr.~Nesterenko and Dr.~Chiba for fruitful discussions.
The computational calculations of this work were performed by using the
supercomputer in the Yukawa Institute for theoretical physics, Kyoto University. 
This work was supported by 
JSPS KAKENHI Grant Nos. 18J20926, 18K03617, and 18H05407.

\bibliographystyle{ptephy}
\bibliography{reference-IMANUM} 

\begin{thebibliography}{10}

\bibitem{1402-4896-2013-T152-014012}
T~Aumann and T~Nakamura, Physica Scripta, {\bf 2013}, 014012 (2013).

\bibitem{Bracco:2015hca}
A.~Bracco, F.~C.~L. Crespi, and E.~G. Lanza, Eur. Phys. J., {\bf A51}, 99
  (2015).

\bibitem{Bracco:2019gza}
A.~Bracco, E.~G. Lanza, and A.~Tamii, Prog. Part. Nucl. Phys., {\bf 106},
  360--433 (2019).

\bibitem{Paar:2007bk}
Nils Paar, Dario Vretenar, Elias Khan, and Gianluca Colo, Rept. Prog. Phys.,
  {\bf 70}, 691--794 (2007).

\bibitem{Harakeh:1981zz}
M.~N. Harakeh and A.~E.~L. Dieperink, Phys. Rev., {\bf C23}, 2329--2334 (1981).

\bibitem{Poelhekken:1992gvp}
T.~D. Poelhekken, S.~K.~B. Hesmondhalgh, H.~J. Hofmann, A.~van~der Woude, and
  M.~N. Harakeh, Phys. Lett., {\bf B278}, 423--427 (1992).

\bibitem{Youngblood:1999zz}
D.~H. Youngblood, Y.~W. Lui, and H.~L. Clark, Phys. Rev., {\bf C60}, 014304
  (1999).

\bibitem{John:2003ke}
Bency John, Y.~Tokimoto, Y.~W. Lui, H.~L. Clark, X.~Chen, and D.~H. Youngblood,
  Phys. Rev., {\bf C68}, 014305 (2003).

\bibitem{PhysRevC.43.2127}
M.~Gai, M.~Ruscev, D.~A. Bromley, and J.~W. Olness, Phys. Rev. C, {\bf 43},
  2127--2139 (1991).

\bibitem{Manley:1991zz}
D.~M. Manley et~al., Phys. Rev., {\bf C43}, 2147--2161 (1991).

\bibitem{Nakatsuka:2017dhs}
N.~Nakatsuka et~al., Phys. Lett., {\bf B768}, 387--392 (2017).

\bibitem{Gibelin:2007fda}
J.~Gibelin et~al. (2007).

\bibitem{Gibelin:2008zz}
J.~Gibelin et~al., Phys. Rev. Lett., {\bf 101}, 212503 (2008).

\bibitem{Brown:2000pd}
B.~Alex Brown, Phys. Rev. Lett., {\bf 85}, 5296--5299 (2000).

\bibitem{Piekarewicz2014}
J.~Piekarewicz, The European Physical JOURNAL A, {\bf 50}, 25 (2014).

\bibitem{Colo2014}
G.~Col{\`o}, U.~Garg, and H.~Sagawa, The European Physical JOURNAL A, {\bf 50},
  26 (2014).

\bibitem{Piekarewicz:2006ip}
J.~Piekarewicz, Phys. Rev., {\bf C73}, 044325 (2006).

\bibitem{Inakura:2011mv}
Tsunenori Inakura, Takashi Nakatsukasa, and Kazuhiro Yabana, Phys. Rev., {\bf
  C84}, 021302 (2011).

\bibitem{Piekarewicz:2012pp}
J.~Piekarewicz, B.~K. Agrawal, G.~Colo, W.~Nazarewicz, N.~Paar, P.~G. Reinhard,
  X.~Roca-Maza, and D.~Vretenar, Phys. Rev., {\bf C85}, 041302 (2012).

\bibitem{Tamii:2011pv}
A.~Tamii et~al., Phys. Rev. Lett., {\bf 107}, 062502 (2011).

\bibitem{Birkhan:2016qkr}
J.~Birkhan et~al., Phys. Rev. Lett., {\bf 118}, 252501 (2017).

\bibitem{Goriely:1998utv}
S.~Goriely, Phys. Lett., {\bf B436}, 10--18 (1998).

\bibitem{Goriely:2004qb}
S.~Goriely, P.~Demetriou, H.~Th. Janka, J.~M. Pearson, and M.~Samyn, Nucl.
  Phys., {\bf A758}, 587--594 (2005).

\bibitem{Tonchev:2017ily}
A.~P. Tonchev et~al., Phys. Lett., {\bf B773}, 20--25 (2017).

\bibitem{Ikeda:pygmy}
Kiyomi Ikeda, INS Report JHP-7 (in Japan) (1988).

\bibitem{Mohan:1971tz}
Radhe Mohan, M.~Danos, and L.~C. Biedenharn, Phys. Rev., {\bf C3}, 1740--1749
  (1971).

\bibitem{10.1143/PTP.83.180}
Yasuyuki Suzuki, Kiyomi Ikeda, and Hiroshi Sato, Prog. Theor. Phys., {\bf 83},
  180--184 (1990).

\bibitem{VanIsacker:1992zz}
P.~Van~Isacker, M.~A. Nagarajan, and D.~D. Warner, Phys. Rev., {\bf C45},
  R13--R16 (1992).

\bibitem{Dubovik:toroidal}
V.~M. Dubovik and Cheshkov~A. A., Sov. J. Part. Nucl., {\bf 5}, 318 (1975).

\bibitem{Semenko:toroidal}
S.~F. Semenko, Sov. J. Nucl. Phys., {\bf 34}, 356 (1981).

\bibitem{Vretenar:2001te}
D.~Vretenar, N.~Paar, and P.~Ring, Phys. Rev., {\bf C65}, 021301 (2002).

\bibitem{Ryezayeva:2002zz}
N.~Ryezayeva, T.~Hartmann, Y.~Kalmykov, H.~Lenske, P.~von Neumann-Cosel,
  V.~{\relax Yu}. Ponomarev, A.~Richter, A.~Shevchenko, S.~Volz, and
  J.~Wambach, Phys. Rev. Lett., {\bf 89}, 272502 (2002).

\bibitem{0954-3899-29-4-312}
J~Kvasil, N~Lo Iudice, Ch~Stoyanov, and P~Alexa, J. Phys. G: Nucl. Part. Phys.,
  {\bf 29}, 753 (2003).

\bibitem{Reinhard:2013xqa}
P.~G. Reinhard, V.~O. Nesterenko, A.~Repko, and J.~Kvasil, Phys. Rev., {\bf
  C89}, 024321 (2014).

\bibitem{Chiba:2015khu}
Y.~Chiba, M.~Kimura, and Y.~Taniguchi, Phys. Rev., {\bf C93}, 034319 (2016).

\bibitem{Kvasil:2013yca}
J.~Kvasil, V.~O. Nesterenko, W.~Kleinig, and P.~G. Reinhard, Phys. Scripta,
  {\bf 89}, 054023 (2014).

\bibitem{Nesterenko:2016qiw}
V.~O. Nesterenko, J.~Kvasil, A.~Repko, W.~Kleinig, and P.~G. Reinhard, Phys.
  Atom. Nucl., {\bf 79}, 842--850 (2016).

\bibitem{Nesterenko:2017rcc}
V.~O. Nesterenko, A.~Repko, J.~Kvasil, and P.~G. Reinhard, Phys. Rev. Lett.,
  {\bf 120}, 182501 (2018).

\bibitem{Shikata:2019wdx}
Yuki Shikata, Yoshiko Kanada-En'yo, and Hiroyuki Morita, PTEP, {\bf 2019},
  063D01 (2019).

\bibitem{Kanada-Enyo:2001yji}
Yoshiko Kanada-En'yo and Hisashi Horiuchi, Prog. Theor. Phys. Suppl., {\bf
  142}, 205 (2001).

\bibitem{KANADAENYO2003497}
Yoshiko Kanada-En'yo, Masaaki Kimura, and Hisashi Horiuchi, Comptes Rendus
  Physique, {\bf 4}, 497--520 (2003).

\bibitem{Kimura:2016fce}
M.~Kimura, T.~Suhara, and Y.~Kanada-En'yo, Eur. Phys. J., {\bf A52}, 373
  (2016).

\bibitem{Dote:1997zz}
Akinobu Dote, Hisashi Horiuchi, and Yoshiko Kanada-En'yo, Phys. Rev., {\bf
  C56}, 1844--1854 (1997).

\bibitem{10.1143/PTP.106.1153}
Masaaki Kimura, Yoshio Sugawa, and Hisashi Horiuchi, Progress of Theoretical
  Physics, {\bf 106}, 1153--1177 (2001).

\bibitem{Suhara:2009jb}
Tadahiro Suhara and Yoshiko Kanada-En'yo, Prog. Theor. Phys., {\bf 123},
  303--325 (2010).

\bibitem{Fujimura:1999zz}
K.~Fujimura, D.~Baye, P.~Descouvemont, Y.~Suzuki, and K.~Varga, Phys. Rev.,
  {\bf C59}, 817--825 (1999).

\bibitem{Descouvemont:2002mnw}
P.~Descouvemont, Nucl. Phys., {\bf A699}, 463--478 (2002).

\bibitem{vonOertzen1996}
W.~von Oertzen, Z. Phys. A, {\bf 354}, 37--43 (1996).

\bibitem{Itagaki:2000nn}
N.~Itagaki, S.~Okabe, and K.~Ikeda, Phys. Rev., {\bf C62}, 034301 (2000).

\bibitem{Kanada-Enyo:1999bsw}
Y.~Kanada-En'yo, H.~Horiuchi, and A.~Dote, Phys. Rev., {\bf C60}, 064304
  (1999).

\bibitem{Kanada-Enyo:2015knx}
Yoshiko Kanada-En'yo, Phys. Rev., {\bf C93}, 024322 (2016).

\bibitem{Kanada-Enyo:1998onp}
Y.~Kanada-En'yo, Phys. Rev. Lett., {\bf 81}, 5291 (1998).

\bibitem{Hill:1952jb}
David~Lawrence Hill and John~Archibald Wheeler, Phys. Rev., {\bf 89},
  1102--1145 (1953).

\bibitem{Griffin:1957zza}
James~J. Griffin and John~A. Wheeler, Phys. Rev., {\bf 108}, 311--327 (1957).

\bibitem{Kanada-Enyo:2019hrm}
Yoshiko Kanada-En'yo and Yuki Shikata, Phys. Rev., {\bf C100}, 014301 (2019).

\bibitem{Kanada-Enyo:2011ldi}
Yoshiko Kanada-En'yo and Tadahiro Suhara, Phys. Rev., {\bf C85}, 024303 (2012).

\bibitem{Kanada-Enyo:2017ers}
Yoshiko Kanada-En'yo, Phys. Rev., {\bf C96}, 034306 (2017).

\bibitem{Volkov:1965zz}
A.~Volkov, Nucl. Phys., {\bf 74}, 33--58 (1965).

\bibitem{Yamaguchi:1979hf}
N.~Yamaguchi, T.~Kasahara, S.~Nagata, and Y.~Akaishi, Prog. Theor. Phys., {\bf
  62}, 1018--1034 (1979).

\bibitem{Tamagaki:1968zz}
R.~Tamagaki, Prog. Theor. Phys., {\bf 39}, 91--107 (1968).

\bibitem{Kanada-Enyo:2006rjf}
Y.~Kanada-En'yo, Prog. Theor. Phys., {\bf 117}, 655--680 (2007).

\bibitem{PhysRevC.80.034318}
C.~M. Mattoon, F.~Sarazin, C.~Andreoiu, A.~N. Andreyev, R.~A.~E. Austin, G.~C.
  Ball, R.~S. Chakrawarthy, D.~Cross, E.~S. Cunningham, J.~Daoud, P.~E.
  Garrett, G.~F. Grinyer, G.~Hackman, D.~Melconian, C.~Morton, C.~Pearson,
  J.~J. Ressler, J.~Schwarzenberg, M.~B. Smith, and C.~E. Svensson, Phys. Rev.
  C, {\bf 80}, 034318 (2009).

\bibitem{Kanada-Enyo:2017uzz}
Yoshiko Kanada-En'yo and Yuki Shikata, Phys. Rev., {\bf C95}, 064319 (2017).

\bibitem{Nesterenko:2019dnt}
V.~O. Nesterenko, A.~Repko, J.~Kvasil, and P.~G. Reinhard, Phys. Rev., {\bf
  C100}, 064302 (2019).

\end{thebibliography}

\end{document}